\documentclass[aps,twocolumn,showpacs,pra]{revtex4}
\usepackage{graphicx,color,amsmath,amssymb,amsbsy,subfigure,hyperref,bbm,times,txfonts}
\begin{document}

\newcommand{\ket}[1]{|#1\rangle}
\newcommand{\bra}[1]{\langle#1|}
\newcommand{\ketbra}[1]{| #1\rangle\!\langle #1 |}
\newcommand{\kebra}[2]{| #1\rangle\!\langle #2 |}
\providecommand{\tr}[1]{\text{tr}\left[#1\right]}
\providecommand{\tra}[1]{\text{tr}_A\left[#1\right]}
\providecommand{\trb}[1]{\text{tr}_B\left[#1\right]}
\providecommand{\abs}[1]{\left|#1\right|}
\providecommand{\sprod}[2]{\langle#1|#2\rangle}
\providecommand{\expect}[2]{\bra{#2} #1 \ket{#2}}

\title{Quantum resources for hybrid communication via qubit-oscillator states}
\author{Tommaso Tufarelli$^1$, Davide Girolami$^2$, Ruggero Vasile$^2$, Sougato Bose$^3$ and Gerardo Adesso$^2$}
\affiliation{$^1$QOLS, Blackett Laboratory, Imperial College London, London SW7 2BW, United Kingdom \\
$^2\hbox{School of Mathematical Sciences, The University of Nottingham, University Park, Nottingham NG7 2RD, United Kingdom}$ \\
$^3$Department of Physics and Astronomy, University College London, Gower Street, London WC1E 6BT, United Kingdom}

\begin{abstract}
We investigate a family of qubit-oscillator states as resources for hybrid quantum communication. They result from a mechanism of qubit-controlled displacement on the oscillator. For large displacements, we obtain analytical formulas for  entanglement and other nonclassical correlations, such as entropic and geometric discord, in those states.
We design two protocols for quantum communication using the considered resource states, a hybrid teleportation and a hybrid remote state preparation. The latter, in its standard formulation, is shown to have a performance limited by the initial mixedness of the oscillator, echoing the behaviour of the geometric discord. If one includes a further optimization over non-unitary correcting operations performed by the receiver, the performance is improved to match that of teleportation, which is directly linked to the amount of entanglement.  Both protocols can then approach perfect efficiency even if the oscillator is originally highly thermal. We discuss the critical implications of these findings for the interpretation  of general quantum correlations.
\end{abstract}


\maketitle

\section{Introduction}
Understanding, identifying and exploiting ``quantumness'' in composite systems represent essential steps to grasp the fundamental implications of quantum theory \cite{zurekrmp}, and are of particular relevance in the race for efficient information and communication technology applications defying the classical boundaries \cite{nielsenchuang}. Nonlocality and entanglement are clearcut signatures of nonclassicality \cite{entanglementreview,nonlocalityreview}. However, it has recently been acknowledged that, focusing on aspects of {\it correlations} among quantum systems, manifestations of quantumness can exist even in absence of entanglement \cite{discord}. Quantum discord and related measures of general quantum correlations are receiving widespread attention \cite{merali} as they promise to enable a supraclassical speedup in computational frameworks where entanglement is not robust enough to endure \cite{dqc1}. Some protocols have been identified that appear to take advantage of discord-like correlations, rather than entanglement, for their functionality \cite{review}; still, basic questions about the interpretation of discord remain unanswered.

Some steps have been undertaken to understand the interplay between entanglement and general quantum correlations \cite{review,beyond}. A hierarchic relation  is proven for two qubits \cite{interplay}, involving the negativity ${\cal N}$ \cite{negativity} as an entanglement measure, and the so-called geometric discord ${\cal D}_G$ \cite{dakic} as a nonclassicality indicator: $\sqrt{{\cal D}_G} \geq {\cal N}$. Less is known about the structure of general quantum correlations in high-dimensional systems, with the exception of continuous variable Gaussian states \cite{adessodatta,giordaparis,gamid,gaussgd}.

\begin{figure}[t]
\includegraphics[width=8.5cm]{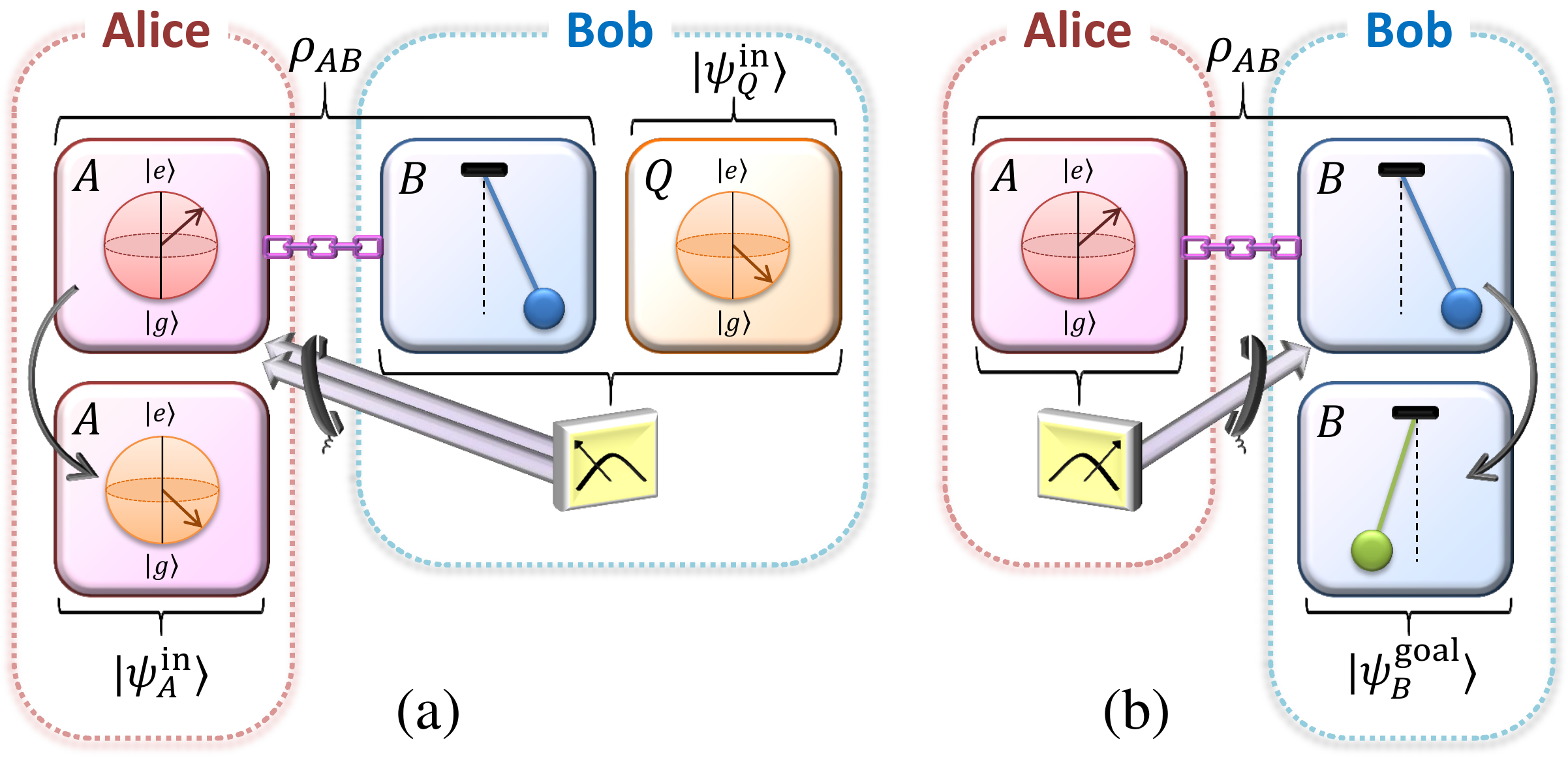}
\caption{(Color online). Schemes for hybrid teleportation (a), and hybrid remote state preparation (b), via the shared resource states $\rho_{AB}$ of a qubit $A$ and an oscillator $B$. Details are provided in the text.}
\label{fische}
\end{figure}

Here we study hybrid bipartite systems comprising a two-level system $A$ (qubit) and a harmonic oscillator $B$ (qumode). We consider a class of states $\rho_{AB}$ in which correlations are induced by the action of qubit-controlled displacements on the oscillator \cite{tufo1}. These states can be implemented in several setups \cite{tufophd} and the required interaction can be exploited for universal quantum computation \cite{qubus}.

The main aim of this paper is construct protocols for quantum communication using these resource states, and analyze their performance in connection with the contents of different types of correlations in the states $\rho_{AB}$.

 We define two state-transfer protocols based on the shared states $\rho_{AB}$, see Fig.~\ref{fische}. One is a hybrid teleportation scheme \cite{telep} where Bob can teleport an unknown qubit state $\ket{\psi^{\rm in}_Q}$ to Alice, and whose fidelity is proven to approach unity for a particular subclass of resource states. The second is a hybrid remote state preparation protocol \cite{originalrsp} where Alice can measure the qubit to remotely prepare Bob's oscillator in some (known to Alice) state $\ket{\psi_B^{\rm goal}}$. In this case, if the receiver Bob can only perform unitary corrections, the fidelity is bounded in general by the initial purity of the oscillator \cite{footnotemixedqub,chaves,dakicrsp}; however, if we allow Bob to perform non-unitary corrections, the figure of merit can increase and match the one associated to teleportation.

We thus analyze in detail the nature of correlations in the states $\rho_{AB}$ and their role for the performance of the different protocols. We obtain analytical formulas for negativity ${\cal N}$ \cite{negativity}, entropic quantum discord ${\cal D}_Z$ \cite{discord}, and geometric discord ${\cal D}_G$ \cite{dakic} of the states $\rho_{AB}$. We find that such states can be maximally entangled and maximally discordant in the limit of large displacements, while their geometric discord is limited by the initial purity of the oscillator and can be thus arbitrarily small.  We argue that this is a direct consequence of the particular geometry of the state-space induced by the Hilbert-Schmidt norm, which enters the definition of ${\cal D}_G$ \cite{dakic}, leading us to conclude that ${\cal D}_G$ cannot be regarded, in general, as a proper `measure' of nonclassical correlations, in agreement with the conclusions of Ref.~\cite{pianicomment}.

This work provides three main advances. First, from a practical viewpoint, it presents workable protocols for hybrid quantum communication, which can be useful as building blocks in any light-matter interfaced implementation of quantum information processing \cite{qinternet,hybridfuru,qubus}. Second, from a technical viewpoint, it provides useful methods for the analytical evaluation of correlation quantifiers in bipartite systems with subsystems of different dimensionality \cite{farlocco}. Third, from a physical perspective, it exposes the need for a mathematically sound and physically meaningful approach to delve into the nature and structure of general nonclassical correlations in quantum states.

The organization of the paper is as follows. In Sec.~\ref{secstates} we define the studied model and calculate analytically various measures of nonclassical correlations for the states $\rho_{AB}$. In Sec.~\ref{secprotocols} we define protocols for teleportation and remote state preparation using the shared resource states $\rho_{AB}$ and calculate their fidelity, relating it to the correlations present in the states. In Sec.~\ref{secconcl} we draw our conclusions. Some technical derivations are deferred to Appendixes.

\section{The states and their correlations.}\label{secstates}
We consider a family of qubit-oscillator states $\rho_{AB}=U_{AB}(\beta)\big(\rho_A^0 \otimes \rho_B^0\big)U^\dagger_{AB}(\beta)$
obtained by applying the unitary $U_{AB}(\beta)=D(\sigma_3 \beta)$ to an initially uncorrelated general state $\rho_A^0 \otimes \rho_B^0$. This interaction induces a qubit-controlled displacement on the oscillator, that is, a displacement where the sign of the parameter $\pm \beta$ is determined by the eigenvalue of the qubit Pauli operator $\sigma_3$ \cite{tufophd,tufo1}. This corresponds e.g.~to the evolution of the joint system via a coupling Hamiltonian $H \propto \sigma_3(b+b^\dagger)$, which can be realized experimentally in a number of setups \cite{tufophd,exprefs}. This type of interaction has relevant applications for the state reconstruction of oscillator networks probed by a single qubit \cite{tufo23} and for quantum computation based on light-matter interfaces \cite{qubus}. We describe the  state of qubit $A$ before the interaction as $\rho_A^0 = {{p \ \ \  r} \choose {\ r^\ast \ 1-p}}$ in the standard basis $\{\ket{e},\ket{g}\}$, with $0 \le p \le 1$ and $|r|^2 \leq p(1-p)$.
The states of the hybrid system after the interaction take the form
\begin{eqnarray}
\rho_{AB}&=&p\kebra{e}{e}\otimes D(\beta)\rho_B^0 D^\dagger(\beta)+(1-p)\kebra{g}{g}\otimes D^\dagger(\beta)\rho_B^0 D(\beta)\nonumber\\
&+&r\kebra{e}{g}\otimes D(\beta)\rho_B^0 D(\beta)+r^*\kebra{g}{e}\otimes D^\dagger(\beta)\rho_B^0 D^\dagger(\beta).\label{states}
\end{eqnarray}
We focus on the regime of large displacements $|\beta| \rightarrow \infty$, which in practice means $|\beta|$ large enough such that the overlap between the two phase-space domains, associated to $\rho_B^0$ displaced by $\beta$, and to $\rho_B^0$ displaced by $-\beta$, becomes negligible.

Let us calculate the correlations in the states of Eq.~(\ref{states}). Nonlocality properties have been studied in \cite{mauro}.

Entanglement can be quantified by the (normalized) negativity \cite{negativity} ${\cal N}(\rho_{AB}) =  (\|\rho_{AB}^{{\sf T}_A}\|_1 -1)$, where $\|M\|_1=\text{Tr}|M|$ denotes the trace norm and $\rho_{AB}^{{\sf T}_A}$ is obtained by partial transposition with respect to the qubit only \cite{ppt}. We find the following rigorous result, whose proof is provided in Appendix~\ref{secappa}
\begin{equation}\label{n}
\lim_{|\beta| \rightarrow \infty} {\cal N}(\rho_{AB}) = 2|r|\,.
\end{equation}
Independently of the initial state of the oscillator (namely, no matter how thermal it is), one can always find a $|\beta|$ large enough such that the qubit-oscillator states are asymptotically maximally entangled, when the qubit is initially in a pure equatorial state, $p=|r|=1/2$. Then the state $\rho_{AB}$ reproduces a proper Schr\"odinger cat state, if one interprets the qubit as the `microscopic' degree of freedom and the qumode as the `macroscopic' one \cite{gatto}.

\begin{figure}[t]
\includegraphics[width=8cm]{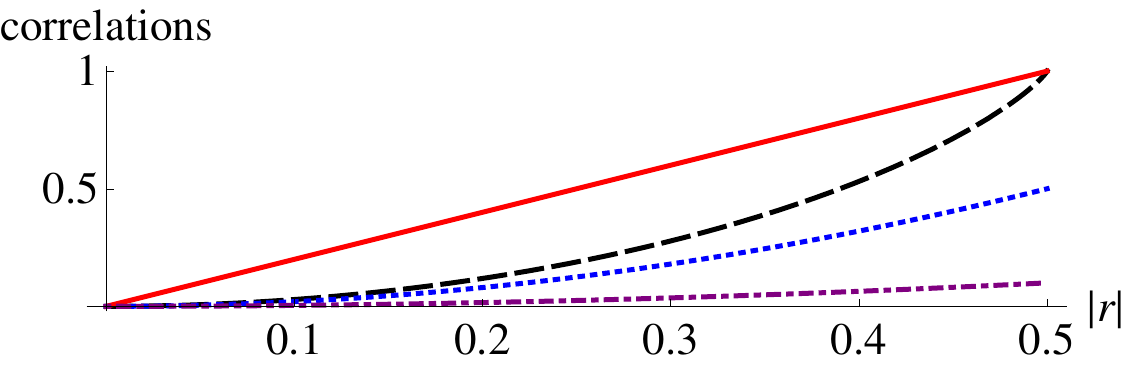}
\caption{(Color online). Plot of negativity [Eq.~\ref{n}] (solid line), lower bound on  entropic discord [Eq.~(\ref{dzdig})] (dashed line), and geometric discord [Eq.~\ref{dg}] for $\mu^0_B=0.5,\ 0.1$ (dotted and dot-dashed line, respectively) calculated for the states $\rho_{AB}$ with $p=1/2$, as a function  of $|r|$. The plotted quantities are dimensionless and correspond to  $|\beta|\rightarrow \infty$.}
\label{figtotcor}
\end{figure}

Let us now consider more general types of nonclassical correlations. The geometric discord ${\cal D}_G(\rho_{AB})$ quantifies how far (in Hilbert-Schmidt norm) a bipartite quantum state is from the set of classical-quantum states \cite{dakic}. It can be measured experimentally with direct, non-tomographic methods \cite{q,cinesidg,passantenew}.
Given a quantum state $\rho_{AB}$ of a $\mathbbm{C}^2 \otimes \mathbbm{C}^d$ system, with $A$ being a qubit and $B$ being an arbitrary (finite or infinite) $d$-dimensional system, the normalized geometric discord ${\cal D}_G(\rho_{AB})$  is defined as \cite{dakic,luofu}
${\cal D}_G(\rho_{AB}) = 2 \inf_{\Pi_A} \|\rho_{AB} -  \Pi_A(\rho_{AB})\|_2^2$, where the infimum is over all von Neumann measurements $\Pi_A \equiv \{\Pi_A^k\}$ on the qubit $A$, with $\Pi_A(\rho_{AB})=\sum_k(\Pi_A^k \otimes \mathbbm{1}_B) \rho_{AB} (\Pi_A^k \otimes \mathbbm{1}_B)$, and $\|M\|_2=\sqrt{\text{Tr}(M M^\dagger)}$ is the Hilbert-Schmidt norm.
In Appendix~\ref{secappb}, we provide a useful and compact analytical framework for the calculation of ${\cal D}_G$ in $2 \times d$ systems including the case $d=\infty$, which we need here. We obtain
\begin{equation}\label{dg}
\lim_{|\beta| \rightarrow \infty} {\cal D}_G(\rho_{AB}) = 4 \mu_B^0  |r|^2 \,.
\end{equation}
The geometric discord can be made arbitrarily small by decreasing the initial purity $\mu_B^0 \equiv \text{tr}_B[({\rho_B^0})^2]$ of the oscillator (see Fig.~\ref{figtotcor}). This result shows that the conjectured ordering relation $\sqrt{{\cal D}_G} \geq {\cal N}$, which holds for all two-qubit states \cite{interplay}, is violated when at least one subsystem of a bipartite system has large dimension, as in our case. For the states $\rho_{AB}$, the ordering is actually {\it reversed}, revealing the quirky situation of states with possibly maximum entanglement yet asymptotically vanishing (for $\mu_B^0 \rightarrow 0$)  geometric discord.

We then ask: are those maximally entangled states infinitesimally close to the classical-quantum border in other metrics?
To address the question, we evaluate the conventional entropic quantum discord ${\cal D}_Z$ \cite{discord} on the states $\rho_{AB}$. We recall the definition, ${\cal D}_Z(\rho_{AB}) = \inf_{\Pi_A} \{{\cal I}(\rho_{AB})-{\cal I}[\Pi_A(\rho_{AB})]\}$, where ${\cal I}$ denotes the quantum mutual information.
We observe that, by means of a local operation on Bob's side, one can map $\rho_{AB}$ onto an effective two-qubit state $\tilde{\rho}_{AB}$. Entanglement measures \cite{entanglementreview} as well as the discord ${\cal D}_Z$ with measurements on $A$ \cite{streltsov}  (but crucially not ${\cal D}_G$ \cite{newdgnonmono})
 are monotonic under such operations on $B$, so that by evaluating those correlation measures on $\tilde{\rho}_{AB}$ one obtains lower bounds to the corresponding measures for $\rho_{AB}$ \cite{ivettebose,lukt,farlocco}. In particular, as proven in Appendix~\ref{secdigapp}, Bob can choose two orthonormal vectors $\ket{\tilde e},\ket{\tilde g}$ and design a local operation which, in the limit of large displacements, converts the state $\rho_{AB}$ to the `digitalized' $\tilde{\rho}_{AB}^{\rm dig} = p \ketbra{e \tilde e} + r \kebra{e\tilde e}{g\tilde g} + r^\ast \kebra{g \tilde g}{e \tilde e} + (1-p) \ketbra{g \tilde g}$, without resorting to post-selection. This effective two-qubit state can be achieved independently of the initial purity of $B$.
The entropic discord in the digitalized state reads
\begin{eqnarray}\label{dzdig}
{\cal D}_Z(\tilde{\rho}_{AB}^{\rm dig})&=&-p \log_2 (p) -(1-p) \log_2(1-p)  \\ &+& \hbox{$\frac12\left[ \log_2\big(p(1-p)-|r|^2\big)+\zeta\log_2\left(\frac{1+\zeta}{1-\zeta}\right)\right]$}, \nonumber
\end{eqnarray}
with $\zeta=\sqrt{(1-2p)^2+4|r|^2}$.
 In the case $p=|r|=1/2$, the entropic discord ${\cal D}_Z(\tilde{\rho}^{\rm dig}_{AB})$ converges to $1$ like the entanglement, implying ${\cal D}_Z({\rho}_{AB})\rightarrow 1$ in the original states of Eq.~(\ref{states}) as well.

This demonstrates that the Hilbert-Schmidt norm is generally unsuitable for defining quantitative distance-based measures of correlations---as recognized for entanglement in \cite{ozawa} and very recently for geometric discord in \cite{pianicomment}---with its deficiency being even more critically exposed in systems with large dimension.  We can then reassess some findings in the recent literature, such as ${\cal D}_G$ failing to capture the resource power of the discrete quantum computation with one bit \cite{dakic,passantenew}, as evidences of this deficiency. Nevertheless, when the purity of the states is fixed, and/or when the dimension of the Hilbert space is small enough (e.g.~for two qubits), the geometric discord returns a reliable quantification of nonclassical correlations \cite{interplay,davidonzo,dakicrsp}. In general, given its computability and experimental accessibility \cite{cinesidg,q}, it can still play a useful role if one correctly regards it not as a measure by itself, but as a valid {\it lower bound} to regular measures of nonclassical correlations like the relative entropy of discord \cite{watrousnotes,req} (alias one-way deficit \cite{onewaydeficit}). Such a bound becomes  looser with increasing dimension, as evidenced by our analysis, and as emerged in the study of continuous variable Gaussian states \cite{gaussgd}.

\section{Hybrid quantum communication protocols}\label{secprotocols}

We now question whether the different aspects of correlations identified in the states of Eq.~(\ref{states}) (see Fig.~\ref{figtotcor} for a comparison) can be endowed with operational meanings.
We design two hybrid quantum communication protocols which employ the family of states $\rho_{AB}$, with Alice operating the qubit $A$ and Bob operating the oscillator $B$, as shared resources.

\subsection{Hybrid teleportation}\label{secTele}
\begin{figure}[t]
\includegraphics[width=8.5cm]{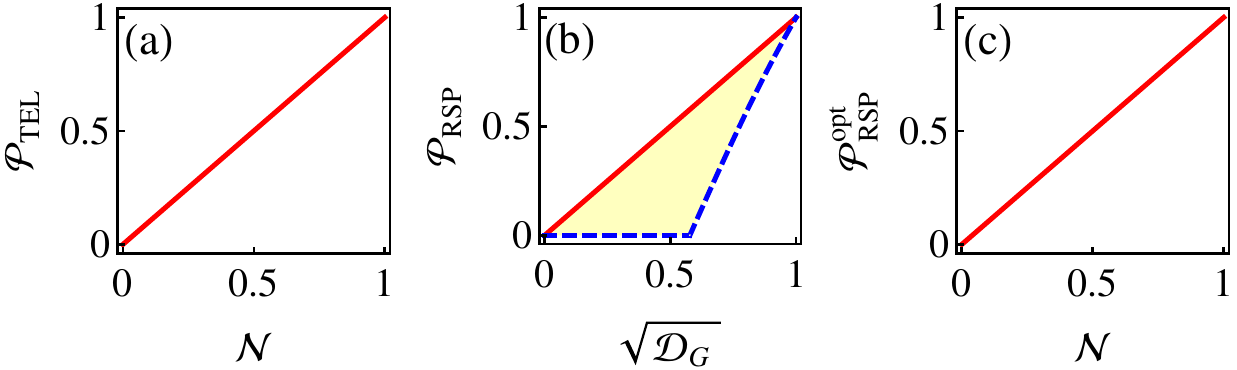}
\caption{(Color online). Payoffs for hybrid teleportation (a), and for hybrid remote state preparation with unitary corrections (b) and optimized with non-unitary corrections (c), plotted versus the correlations in the shared states, measured by negativity in panels (a),(c) and by square root of geometric discord in panel (b); see text for details on the boundary curves in (b).
The plotted quantities are dimensionless and correspond to  $|\beta|\rightarrow \infty$.}
\label{payoffighi}
\end{figure}

Alice and Bob share a state $\rho_{AB}$ of the form (\ref{states}) [See Fig.~\ref{fische}(a) for reference]. Bob (the sender) wishes to teleport \cite{telep} an unknown state $\ket{\psi^{{\rm in}}_Q} = \eta\ket{e} + \gamma\ket{g}$ of an input qubit $Q$ to Alice (the receiver). Bob then makes a joint measurement on the input qubit $Q$ and the oscillator $B$, communicates the outcome to Alice, who can then implement a correction on her qubit $A$. The final state $\rho_A^{{\rm out}}$ of $A$ can be shown (see below) to have a fidelity ${\cal F}_{\textsc{tel}}=\bra{\psi^{{\rm in}}_Q}\rho_A^{{\rm out}}\ket{\psi^{{\rm in}}_Q}$, averaged over the uniform  distribution of the input state, given by
\begin{equation}\label{fidtelep}
\lim_{|\beta| \rightarrow \infty} \bar{{\cal F}}_{\textsc{tel}} = \tfrac23\big(1+|r|\big)\,.
\end{equation}
Notice that for any $|r|>0$ the average fidelity $\bar{{\cal F}}_{\textsc{tel}}$ exceeds the classical benchmark achievable by measure-and-prepare schemes, ${\cal F}^{\rm cl}_{\textsc{tel}} = 2/3$ \cite{telebenchmark}. We can then define a {\it payoff} \begin{equation}{\cal P}_{\textsc{tel}} = (1-{\cal F}^{\rm cl}_{\textsc{tel}})^{-1}\max\{0, \bar{{\cal F}}_{\textsc{tel}} - {\cal F}^{\rm cl}_{\textsc{tel}}\}\end{equation} quantifying the better-than-classical performance of the teleportation protocol. It is immediate to see that in the limit of large displacements
\begin{equation}\label{payofftelep}
{\cal P}_{\textsc{tel}} = {\cal N}(\rho_{AB})=2|r|\,.
\end{equation}
This shows that entanglement, in the form of negativity, is clearly the resource for this protocol [see Fig.~\ref{payoffighi}(a)].

We now provide the explicit steps of the protocol and prove the result announced in Eq.~(\ref{fidtelep}), which can be formalized as follows: For any $\epsilon>0$, the state \eqref{states} can be used for teleportation of a generic qubit state $\ket{\psi_Q^{\rm in}}$  from Bob to Alice, with average fidelity $\bar{{\cal F}}_{\textsc{tel}}\geq (2/3)(1+|r|)-\epsilon$ and success probability $P\geq 1-\epsilon$.

We first choose a cutoff integer $N$ and define $\epsilon_N=\sum_{N+1}^\infty s_n$, where the initial state of the oscillator $B$ in Eq.~(\ref{states}) has been written generically as $\rho_B^0 = \sum_n s_n \ketbra{\psi_n}$.  We can find a lower bound to the fidelity by just assuming that, with probability $\epsilon_N$, the protocol fails. That is, we may use
$\rho_\textrm{AB}'=(1-\epsilon_N)\rho_{AB}^{(N)}+\epsilon_N\rho_{AB}^{\perp}$
where $\rho_{AB}^{(N)}$ is the truncated version of $\rho_\textrm{AB}$ and $\rho_{AB}^\perp$ a state that yields zero teleportation fidelity. From now on we  assume that Alice and Bob share the truncated resource $\rho_{AB}^{(N)}$, and the derived results will hold with probability $P\geq1-\epsilon_N$. One can always choose $N$ large enough so that $\epsilon_N<\epsilon$.

The initial state of the complete $ABQ$ system is
$\rho^{{\rm in}}_{ABQ}=\ketbra{\psi_Q^{\rm in}}\otimes\rho_{AB}^{(N)}$.
The state $\rho_{AB}^{(N)}$ can be expanded in terms of the qubit basis and the $2(N+1)$ states $\{D(\beta)\ket{\psi_n},D^\dagger(\beta)\ket{\psi_n}\}_{n=0,...,N}$. If we choose $|\beta|$ large enough, these states will be effectively orthonormal. Then, Bob performs a hybrid Bell-state measurement corresponding to the following $4(N+1)$ orthonormal vectors, $\ket{\phi_m^\pm}=\tfrac{1}{\sqrt2}\big(\ket e D(\beta)\ket{\psi_m}\pm \ket g D^\dagger(\beta)\ket{\psi_m}\big)$, and $\ket{\xi_m^\pm}=\tfrac{1}{\sqrt2}\big(\ket e D^\dagger(\beta)\ket{\psi_m}\pm \ket g D(\beta)\ket{\psi_m}\big)$. These measurements can be done as follows: Bob performs the disentangling operation $U^\dagger=D^\dagger(\sigma_3\beta)$ on the input qubit $Q$ and his oscillator $B$, then measures the qubit $Q$ in the basis $\{\ket e\pm \ket g\}$, and the oscillator in the effectively orthogonal basis $\{\ket{\psi_m},D(2\beta)\pm D^\dagger(2\beta)]\ket{\psi_m}\}_{m=0}^N$ \cite{noteex}.
Then, upon receiving two classical bits from Bob (as all values of $m$ give the same results), Alice first performs a local phase rotation $\ket e\to\tfrac{|r|}{r}\ket e$ on qubit $A$; further, she may or may not perform the corrections $\ket e\leftrightarrow \ket g$ and $\ket e\to-\ket e$, depending on Bob's outcomes. After tedious calculations, we find that there are only two possible (unnormalized) states that Alice gets,
\begin{eqnarray*}
\rho^{{\rm out} \phi}_A&=& p|\eta|^2\kebra{e}{e}+(1-p)|\gamma|^2\kebra{g}{g}+ |r|\eta\gamma^*\kebra{e}{g}+  |r|\eta^*\gamma\kebra{g}{e},\\
\rho^{{\rm out} \xi}_A&=& (1-p)|\eta|^2\kebra{e}{e}+p|\gamma|^2\kebra{g}{g}+ |r|\eta\gamma^*\kebra{e}{g}+ |r|\eta^*\gamma\kebra{g}{e}.\end{eqnarray*} The input-output fidelity is then given by \begin{equation}{\cal F}_{\textsc{tel}} = \sum_{j=\phi,\xi}\bra{\psi^{{\rm in}}_Q} {\rho}_A^{{\rm out} j} \ket{\psi^{{\rm in}}_Q} =|\eta|^4+|\gamma|^4+4|r| |\eta|^2|\gamma|^2.\end{equation} Averaging $\eta,\gamma$ over the Bloch spere, we obtain Eq.~(\ref{fidtelep}).
%

\subsection{Hybrid remote state preparation}\label{secRSP}
We now describe a different protocol, see Fig.~\ref{fische}(b) for reference.  Alice and Bob share again a state $\rho_{AB}$ of the form (\ref{states}). Without loss of generality, we can assume $r=|r|$ (this is true up to a local unitary on $A$). Alice (the preparer) wishes to remotely prepare \cite{originalrsp} Bob's oscillator in the target superposition state, known to Alice, \begin{equation}\ket{\psi_B^{\rm goal}} = (\ket{\beta}+{\rm e}^{-i \varphi}\ket{\!-\!\beta})/\sqrt{2(1+\cos\varphi{\rm e}^{-2 |\beta|^2})}\,,\end{equation} for some phase $\varphi$. For this purpose, Alice measures qubit $A$ in the basis $\ket{\pm_A}=(\ket{e} \pm {\rm e}^{i \varphi} \ket{g})/\sqrt2$ and classically communicates the one-bit outcome ``$\pm$'', obtained with probability $P_\pm$, to Bob. If Alice obtains ``+'', Bob does nothing, otherwise he applies a $\pi$-phase shift $\Phi_B$ to $B$, which ideally (for large $|\beta|$) maps $\ket{\beta}\rightarrow \ket{\beta}$ and $-\ket{\beta}\rightarrow-\ket{\!-\!\beta}$, and can be implemented in phase space via a combination of displacements and photon subtraction \cite{fiurasekexp}. The qumode $B$ after the correction can be in one of two possible unnormalized states, $\rho_B^{{\rm out} +}= \bra{+_A} \rho_{AB} \ket{+_A}$ and $\rho_B^{{\rm out} -}=\Phi_B\bra{-_A} \rho_{AB} \ket{-_A}\Phi_B^\dagger$.
The fidelity of the protocol, averaged over the distribution of the target state, is then \begin{equation}\bar{\cal F}_{\textsc{rsp}} = (2\pi)^{-1} \int {\rm d} \varphi \bra{\psi^{\rm goal}_B} (\rho_B^{{\rm out} +}+\rho_B^{{\rm out} -}) \ket{\psi^{\rm goal}_B}\,.\end{equation} Clearly the ideal resource for this protocol is obtained for $p=|r|=1/2$ and $\rho_B^0=\kebra{0}{0}$, where $\ket0$ is the ground state of the oscillator. A typical realistic deviation is given by the oscillator being initially in a mixed state. For simplicity, let us focus on $\rho_B^0$ being in general a Gaussian thermal state \cite{quantumoptics} $\rho_B^0=\sum_n s_n \ketbra{n}$, where $s_n =\bar{n}^n/(1+\bar{n})^{n+1}$, with $\bar{n}$ the mean number of thermal excitations. This choice allows us to compute the fidelity analytically (although different choices for $\rho_B^0$ give rise to qualitatively similar results as verifiable numerically), yielding
\begin{equation}\label{fidrsp}
\lim_{|\beta|\rightarrow \infty} \bar{\cal F}_{\textsc{rsp}} = \tfrac{\mu_B^0}{1+\mu_B^0}\big(1+2|r|\big)\,.
\end{equation}
In this case, the fidelity is limited by the initial purity of the oscillator, akin to the geometric discord in Eq.~(\ref{dg}). For remote state preparation, the classical threshold corresponds to Bob preparing a completely random guess (unnormalized) state \cite{originalrsp,dakicrsp} $\rho_B^{\rm cl}=(\kebra{\beta}{\beta}+\kebra{\!-\!\beta}{\!-\!\beta})/2$, yielding \begin{equation}{\cal F}_{\textsc{rsp}}^{\rm cl} = (2\pi)^{-1} \int {\rm d} \varphi \bra{\psi^{\rm goal}_B} \rho_B^{\rm cl} \ket{\psi^{\rm goal}_B} \underset{|\beta|\rightarrow \infty}{\longrightarrow} \frac12\,.\end{equation} Defining again the {\it payoff} as \begin{equation}{\cal P}_{\textsc{rsp}} = (1-{\cal F}^{\rm cl}_{\textsc{rsp}})^{-1}\max\{0, \bar{{\cal F}}_{\textsc{rsp}} - {\cal F}^{\rm cl}_{\textsc{rsp}}\}\end{equation}, we then find in the limit of large displacements
\begin{equation}\label{payoffrsp}
\hbox{$L[{\cal D}_G(\rho_{AB})]$}
\leq {\cal P}_{\textsc{rsp}}\! =\! \max\left\{0,\,\hbox{$\frac{\mu_B^0(1+4|r|)-1}{1+\mu_B^0}$}\right\} \leq\! \hbox{$\sqrt{\!{\cal D}_G(\rho_{AB})}$}\,.
\end{equation}
For a given ${\cal D}_G(\rho_{AB})$, the payoff in the performance of remote state preparation can never exceed $\sqrt{{\cal D}_G}$ [which is reached for $\mu_B^0=1$, solid line in Fig.~\ref{payoffighi}(b)], and admits a lower bound as well,  $L[{\cal D}_G] =\max\{0,\,(3{\cal D}_G-1)/(1+{\cal D}_G)\}$ [dashed curve in Fig.~\ref{payoffighi}(b)]. The latter is tight for $p=|r|=1/2$, when the shared states are {\it maximally entangled}, yet the remote state preparation succeeds with null or limited payoff.
A nonzero payoff implies necessarily a nonzero ${\cal D}_G$. Thus, despite its clear shortcomings \cite{pianicomment}, the geometric discord might still seem to capture the operative performance of qubit-to-oscillator remote state preparation, in analogy with the case of two-qubit resources \cite{dakicrsp}.

A remark is in order. In the previously described hybrid teleportation (Sec.~\ref{secTele}), Bob can optimize his measurement strategy to compensate for the initial mixedness of the oscillator, which therefore does not affect the achievable fidelity. In contrast, in remote state preparation the measurement of Alice's qubit necessarily leaves Bob's oscillator in a mixed state, if its
initial state $\rho^0_B$ was mixed. As the protocol aims at preparing a pure state, it is no surprise that the resource mixedness enters in, and degrades, the figure of merit.
We can address this limitation by extending the conventional remote state preparation primitive \cite{originalrsp}, to allow Bob to perform general completely positive maps rather than just unitary corrections, aiming to improve the fidelity with the goal state \cite{fiuracoso}. In our hybrid case, we observe that if Bob performs the operation which `digitalizes' the resource state as discussed above, then the mixedness of the oscillator is effectively bypassed and we obtain a payoff ${\cal P}^{\rm opt}_{\textsc{rsp}}=2|r|$, equal to the hybrid teleportation payoff and coinciding with the negativity of $\rho_{AB}$ [see Fig.~\ref{payoffighi}(c)]. We now present details of the procedure.

We start with the usual resource state $\rho_{AB}$ given by Eq.~(\ref{states}).
Alice wants to prepare remotely a state of the form $\ket{\psi_B^{\rm goal}}=\tfrac{1}{\sqrt2}(\ket{\tilde e}+{\rm e}^{-i\varphi}\ket{\tilde{g}})$, where $\ket{\tilde e}$ and $\ket{\tilde g}$ are two orthogonal states in Bob's Hilbert space. Alice first rotates her basis such that $r\to|r|$, then performs a projective measurement on her part of the system, using the basis $\ket{\!\pm\!\varphi}=\tfrac{1}{\sqrt2}(\ket e\pm{\rm e}^{-i\varphi}\ket g)$. The reduced density matrix of Bob after the measurement (up to a normalization constant) is given by
\begin{align}
\bra{\pm\varphi}\rho_{AB}\ket{\!\pm\!\varphi}&=pD(\beta)\rho_B^0 D^\dagger(\beta)+(1-p)D^\dagger(\beta)\rho_B^0 D(\beta)+\nonumber\\
&\pm |r|{\rm e}^{i\varphi} D(\beta)\rho_B^0 D(\beta)\pm |r|{\rm e}^{-i\varphi} D^\dagger(\beta)\rho_B^0 D^\dagger(\beta).
\end{align}
In the limit of large displacements, Bob can perform the following local (non-unitary) operation, with probability approaching unity (Note: it is the same `digitalizing' operation detailed in Appendix~\ref{secdigapp}):
\begin{align}
D(\beta)\rho_B^0 D^\dagger(\beta)&\to\kebra{\tilde e}{\tilde e},\nonumber\\
D^\dagger(\beta)\rho_B^0 D(\beta)&\to\kebra{\tilde g}{\tilde g},\label{dito}\\
D(\beta)\rho_B^0 D(\beta)&\to\kebra{\tilde e}{\tilde g}.\nonumber
\end{align}
The state of Bob's mode then becomes
\begin{equation}
\rho_B'=p\kebra{\tilde e}{\tilde e}+(1-p)\kebra{\tilde g}{\tilde g}\pm|r|{\rm e}^{-i\varphi}\kebra{\tilde e}{\tilde g}\pm|r|{\rm e}^{i\varphi}\kebra{\tilde g}{\tilde e},
\end{equation}
which is now properly normalized. Finally, Bob can remove the $\pm$ signs with a further local unitary correction, obtaining eventually the state
\begin{equation}
\rho_B''=p\kebra{\tilde e}{\tilde e}+(1-p)\kebra{\tilde g}{\tilde g}+|r|{\rm e}^{-i\varphi}\kebra{\tilde e}{\tilde g}+|r|{\rm e}^{i\varphi}\kebra{\tilde g}{\tilde e}.
\end{equation}
The fidelity between this output state and the target state is
\begin{equation}
{\cal F}^{\rm opt}_{\textsc{rsp}}=\bra{\psi_B^{\rm goal}}\rho_B''\ket{\psi_B^{\rm goal}}=\frac{1}{2}+|r|,
\end{equation}
which does not depend on the angle and is hence equal to the average fidelity $\bar{{\cal F}}^{\rm opt}_{\textsc{rsp}}$. Recalling that for remote state preparation ${\cal F}^{\rm cl}_{\textsc{rsp}} \rightarrow \frac12$, we finally recover that the payoff, for the optimized scheme incorporating a non-unitary local correction on Bob's side, becomes ${\cal P}^{\rm opt}_{\textsc{rsp}} = 2 |r|$ in the limit of large displacements, as anticipated above.

This suggests that the link between remote state preparation and measures of discord, highlighted for two-qubit systems \cite{chaves,dakicrsp}, might be due to a non-optimized version of the protocol used. In the case of the qubit-oscillator resources described here, such limitation may be relevant where the `digitalizing' operation of Eq.~\eqref{dito} is experimentally challenging to realize, so that one is constrained to unitary corrections only. However, the resulting connection between geometric discord and protocol performance would be solely due to technological limitations, and not to fundamental quantum-mechanical principles. In fact, we showed how properly accounting for an extra freedom to correct for the resource mixedness allows the protocol to reach a performance only dependent on the amount of shared entanglement.

It will be interesting to test how this result is modified once additional decoherence sources in the implementation of the hybrid protocols are considered, to see whether the link with entanglement will persist also for remote state preparation, or some form of discord would emerge as essential operational ingredient.

\section{Concluding remarks}\label{secconcl}
We conclude that, broadly speaking, various indicators of ``quantumness'' in composite systems \cite{entanglementreview,review} can be justified by operational interpretations related to their role in different tasks. To wit, entanglement is a resource for teleportation \cite{telep,telebenchmark} and superdense coding \cite{densecoding} (among all), geometric discord is such for a non-globally-optimized remote state preparation scheme \cite{dakicrsp} (as shown here), entropic discord is interpreted through quantum state merging \cite{operdiscord}, and relative entropy of discord \cite{req,onewaydeficit} is the cost of entanglement distribution via separable states \cite{stronzof}. The list is likely to grow in the future, although care is needed to ensure it does not derails away from physical grounds.

On the practical side, we demonstrated that robust quantum correlations in the form of entropic discord and entanglement can be engineered in coupled qubit-oscillator systems no matter the temperature of the oscillator, provided the interaction generates enough displacement. This indicates that, e.g.~in systems of nano/optomechanical oscillators \cite{vitali} coupled to a two-level probe, there is no need to cool the oscillator down to its ground state, in order for quantum communication to be achieved efficiently. Although this may require generalized (non-unitary) operations that can be challenging to realize with current technology, the question of implementing those operations in a realistic experimental setup can be a source of further interesting research outside the scopes of this paper. This may in the future relax the need for ground state cooling in favor of generalized operations. In fact, while ground state cooling has attracted experimental efforts for a long time, and is hence well developed, our findings support the view that the experimental realization of generalized quantum operations, a comparatively young subject, also may deserve consideration and could have a substantial practical impact for the realistic implementation of quantum technologies.

\acknowledgments{We thank C. Brukner, E. Chitambar, B. Daki\'c, A. Ferraro, B. Garraway, S. Gharibian, V. Giovannetti, T. Nakano, M. G. A. Paris, M. Piani, S. Ragy, V. Vedral and especially M. S. Kim for fruitful discussions. We thank an anonymous referee for having suggested how to improve the remote state preparation protocol. We acknowledge financial support from the Royal Society, the Wolfson Foundation, the University of Nottingham (ECRKTA/2011), the U.K. EPSRC (Grant RDF/BtG/0612b/31), the Qatar National Research Fund (NPRP 4-554-1-084).}

\appendix

\section{Negativity of the qubit-oscillator states $\boldsymbol{\rho_{AB}}$}\label{secappa}
Here we derive Eq.~(\ref{n}), which can be formalized as follows.

\smallskip

\noindent {\bf Theorem 1.} {\it For any $\epsilon>0$, it is possible to find a $\beta \in \mathbbm{C}$ such that \begin{equation}
\label{napp}\mathcal{N}(\rho_{AB})\geq2|r|-\epsilon\,,
 \end{equation} where $\mathcal{N}$ is the negativity \cite{negativity} (twice the modulus of the sum of the negative eigenvalues of $\rho_{AB}^{{\sf T}_A}$).}

\smallskip

\noindent {\it Proof.}
The partial transpose of the state \eqref{states}, with respect to the qubit, is
\begin{align}
\rho_{AB}^{{\sf T}_A}&=p\kebra{e}{e}\otimes D(\beta)\rho^0_B D^\dagger(\beta)+(1-p)\kebra{g}{g}\otimes D^\dagger(\beta)\rho^0_B D(\beta)+\nonumber\\
&+r\kebra{g}{e}\otimes D(\beta)\rho^0_B D(\beta)+r^*\kebra{e}{g}\otimes D^\dagger(\beta)\rho^0_B D^\dagger(\beta).\label{PT}
\end{align}
Let us consider the test states
\begin{equation}
\ket{\phi_m}=\frac{1}{\sqrt2}\left(\ket e D^\dagger(\beta)\ket{\psi_m}-{\rm e}^{i\phi}\ket gD(\beta) \ket{\psi_m}\right),\label{test}
\end{equation}
where $\psi_m$ are the eigenvectors of the initial oscillator state $\rho^0_B=\sum_m s_m \kebra{\psi_m}{\psi_m}$, and $\phi$ is the complex argument of $r$, that is, ${\rm e}^{i\phi}=\tfrac{r}{|r|}$.
The expectation value of the partial transpose on the test states is
\begin{align}
\bra{\phi_m}\rho_{AB}^{{\sf T}_A}\ket{\phi_m}&=-|r|s_m+\frac{p}{2}\expect{D(2\beta)\rho^0_BD^\dagger(2\beta)}{\psi_m} \nonumber \\ &+\frac{1-p}{2}\expect{D^\dagger(2\beta)\rho^0_BD(2\beta)}{\psi_m}.\label{test-exp}
\end{align}
Now, let us make use of the following Lemma (proven below):\\
{\bf Lemma 2.} {\it Given any oscillator density matrix $\rho$ and any oscillator pure state $\psi$, one has
\begin{equation}
\lim_{|\alpha|\rightarrow\infty}\expect{D(\alpha)\rho D^\dagger(\alpha)}{\psi}=0.\label{limit}
\end{equation}
}

Let us then pick a cutoff $N$ such that $\sum_m^Np_m\geq1-\epsilon'$, where $\epsilon'>0$, and, by using the property \eqref{limit}, let us choose $\beta$ such that for any $m\leq N$ one has $\expect{D(2\beta)\rho^0_BD^\dagger(2\beta)}{\psi_m}<\epsilon'$ and also $\expect{D^\dagger(2\beta)\rho^0_BD(2\beta)}{\psi_m}<\epsilon'$. Then, from Eq.~\eqref{test-exp} it follows that
$$
\sum_m^N\bra{\phi_m}\rho_{AB}^{{\sf T}_A}\ket{\phi_m}\leq-|r|+\epsilon'(|r|+N/2).
$$
By choosing $\epsilon'=(\epsilon/2)(|r|+N/2)^{-1}$, and using the fact that the test states are orthonormal, we obtain Eq.~(\ref{napp}).
\hfill $\blacksquare$

\smallskip

\noindent {\it Proof of Lemma 2}.
Given any density operator $\rho$, one has $$\lim_{|\alpha|\rightarrow\infty}\chi_\rho(\alpha)=0,$$ $\chi_\rho(\alpha)=\tr{\rho D(\alpha)}$ being its characteristic function. This is a consequence of the fact that $$\tr{\rho^2}=\pi^{-1}\int{\rm d}^2\alpha |\chi_\rho(\alpha)|^2\leq1.$$
In particular, by taking a projector $\rho=\kebra{\varphi}{\varphi}$, we have
$$
\lim_{|\alpha|\rightarrow\infty}\expect{D(\alpha)}{\varphi}=0.
$$
Then, by substituting $\ket\varphi$ with $\ket{\varphi_1}\pm\ket{\varphi_2},\ket{\varphi_1}\pm i\ket{\varphi_2}$ in the above result, it is easy to show that, for any pair of vectors $\ket{\varphi_1},\ket{\varphi_2}$, one has
\begin{equation}
\lim_{|\alpha|\rightarrow\infty}\bra{\varphi_1}D(\alpha)\ket{\varphi_2}=0.\label{offdiagonal}
\end{equation}
To prove \eqref{limit}, we decompose $\rho=\sum_n q_n\kebra{\varphi_n}{\varphi_n}$, and write
\begin{equation}
\expect{D(\alpha)\rho D^\dagger(\alpha)}{\psi}=\sum_n q_n|\bra{\psi}D(\alpha)\ket{\varphi_n}|^2.
\end{equation}
We now choose $N$ such that $\sum_{N+1}^\infty q_n\leq\epsilon/2$, and thanks to Eq.~\eqref{offdiagonal} we can choose $\alpha$ such that $|\bra{\psi}D(\alpha)\ket{\varphi_n}|^2<\epsilon/2$ for any $n\leq N$. It follows that
\begin{align}
\sum_n &q_n|\bra{\psi}D(\alpha)\ket{\varphi_n}|^2 \nonumber \\ &=\sum_n^N q_n|\bra{\psi}D(\alpha)\ket{\varphi_n}|^2+\sum_{n=N+1}^\infty q_n|\bra{\psi}D(\alpha)\ket{\varphi_n}|^2\nonumber\\
&\leq \frac{\epsilon}{2}\sum_n^N q_n+\sum_{N+1}^\infty q_n\leq \frac{\epsilon}{2}\sum_n^\infty q_n+\frac{\epsilon}{2}=\epsilon
\end{align}
Then, given any $\epsilon>0$, we have found $\alpha$ such that
\begin{equation}
\expect{D(\alpha)\rho D^\dagger(\alpha)}{\psi}<\epsilon,
\end{equation}
which proves Eq.~\eqref{limit}.
\hfill $\blacksquare$

\section{Geometric discord of the qubit-oscillator states $\boldsymbol{\rho_{AB}}$}\label{secappb}

\subsection{Geometric Discord for $2 \otimes \infty$ systems}

Given a quantum state $\rho_{AB}$ of a $\mathbbm{C}^2 \otimes \mathbbm{C}^d$ system, with $A$ being a qubit and $B$ being an arbitrary (finite or infinite) $d$-dimensional system, the normalized geometric discord ${\cal D}_G(\rho_{AB})$  is defined as \cite{dakic,luofu}
$${\cal D}_G(\rho_{AB}) = 2 \inf_{\Pi_A} \|\rho_{AB} -  \Pi_A(\rho_{AB})\|_2^2\,,$$
where the infimum is over all von Neumann measurements $\Pi_A \equiv \{\Pi_A^k\}$ on the qubit $A$, with $\Pi_A(\rho_{AB})=\sum_k(\Pi_A^k \otimes \mathbbm{1}_B) \rho_{AB} (\Pi_A^k \otimes \mathbbm{1}_B)$, and $\|M\|_2=\sqrt{\text{Tr}(M M^\dagger)}$ is the Hilbert-Schmidt norm.

Noting that any projective measurement induces a dephasing of the qubit on some orthonormal basis $\{\ket{c},\ket{c^\perp}\}$, i.e., $\Pi_A(\rho_{AB}) = (\rho_{AB}+U_A \rho_{AB} U^{\dagger}_A)/2$, with $U_A=\ketbra{c} - \ketbra{c^\perp}$ a ``root-of-unity'' operation on $A$ \cite{stellar,gharibian,takano}, we can recast the problem into an optimization over such local unitaries, ${\cal D}_G(\rho_{AB}) =2 \inf_{U_A} \|[\rho_{AB} -  (U_A \otimes \mathbbm{1}_B) \rho_{AB} (U^{\dagger}_A \otimes \mathbbm{1}_B)]/2\|_2^2$. Introducing a generic projector $P_A^{\bf e}=\ketbra{c}=({\bf e} \cdot {\boldsymbol{\sigma}})/2$ on the qubit, where $\boldsymbol{\sigma}=(\mathbbm{1},\vec{\sigma})$ is a four-vector of Pauli matrices, and ${\bf e}=(1,\hat{e})$ with $\hat{e} \in \mathbbm{R}^3$ a unit vector, the geometric discord can be expressed as follows \cite{gharibian},
\begin{equation}
{\cal D}_G(\rho_{AB}) = \inf_{{\bf e}} \big\{4 \text{tr} [\rho_{AB}^2(P_A^{\bf e} \otimes \mathbbm{1}_B)-\rho_{AB}(P_A^{\bf e} \otimes \mathbbm{1}_B)\rho_{AB}(P_A^{\bf e} \otimes \mathbbm{1}_B)]\big\}\,.\label{dgtomin}
\end{equation}

This minimization can be solved in closed form. We define a `partial Fano representation' \cite{fano} of $\rho_{AB}$ by expanding only the qubit $A$ in the Bloch basis, $$\rho_{AB}=\frac12({\bf v} \cdot \boldsymbol{\sigma}),$$ with the four-vector of operators ${\bf v} = \text{tr}_A(\rho_{AB} \boldsymbol{\sigma}) \equiv (v_0, \vec{v})$, where $v_0 \equiv \rho_B$ is the reduced density matrix of $B$.  Like in relativity theory we  use Greek indices to indicate the components $0,1,2,3$ and Roman indices to indicate the `spatial' components $1,2,3$ only.

Let us evaluate each term appearing inside the minimization in Eq.~(\ref{dgtomin}). For the first, we have
\begin{align}
4\tr{\rho_{AB}^2 P_A^{\bf e}}&=\frac{1}{2}\tr{({\bf v}\cdot\boldsymbol{\sigma})^2({\bf e}\cdot\boldsymbol{\sigma})}\nonumber \\&=\frac{1}{2}\trb{v_\mu v_\nu}e_\eta\tra{\sigma_\mu \sigma_\nu \sigma_\eta}, \nonumber
\end{align}
where sum over the repeated indices is understood as in Einstein's convention. To evaluate the second term, we begin by noting that, if $P_A^{\bf e}=({\bf e} \cdot {\boldsymbol{\sigma}})=\kebra{c}{c}$, then
\begin{align}
\tr{\rho_{AB} P_A^{\bf e}\rho_{AB} P_A^{\bf e}}&=\trb{\tra{\rho_{AB}\kebra{c}{c}\rho_{AB}\kebra{c}{c}}}\nonumber\\
&=\trb{\bra c\rho_{AB}\ket c\bra c\rho_{AB}\ket c}=\nonumber\\
&=\trb{\tra{\rho P_{AB}}\tra{\rho P_{AB}}}. \nonumber
\end{align}
Then we have
\begin{align}
4&\tr{\rho_{AB} P_A^{\bf e}\rho_{AB} P_A^{\bf e}}\nonumber \\ \quad &=\frac{1}{4}\trb{\tra{({\bf v}\cdot\boldsymbol{\sigma})({\bf e}\cdot\boldsymbol{\sigma})}\tra{({\bf v}\cdot\boldsymbol{\sigma})({\bf e}\cdot\boldsymbol{\sigma})}}=\nonumber\\
&=\frac{1}{4}\trb{v_\mu v_\nu}e_\eta e_\tau\tra{\sigma_\mu\sigma_\eta}\tra{\sigma_\nu\sigma_\tau}=\nonumber\\
&=\trb{v_\mu v_\nu}e_\mu e_\nu. \nonumber
\end{align}
For the last equality we have used the fact that $\tra{\sigma_\mu\sigma_\eta}=2\delta_{\mu\eta}$. We now define the $4 \times 4$ matrix
$$
\mathcal S_{\mu\nu}=\trb{v_\mu v_\nu}.
$$
Note that the matrix $\mathcal S$ is symmetric, due to the cyclic invariance of the trace.
Then Eq.~(\ref{dgtomin}) can be rewritten as ${\cal D}_G(\rho_{AB}) = \inf_{{\bf e}} T$, with
\begin{align}
T&=\frac{1}{2}\mathcal S_{\mu\nu}e_\eta\tra{\sigma_\mu \sigma_\nu \sigma_\eta}-\mathcal S_{\mu\nu}e_\mu e_\nu=\nonumber\\
&=\frac{1}{2}\mathcal S_{ij}e_\eta\tra{\sigma_i \sigma_j \sigma_\eta}-\mathcal S_{ij}e_i e_j,\label{T-one}
\end{align}
where we have used the fact that the contributions with $\mu=0$ or $\nu=0$ vanish (this is easy to check in the above formula noting that $e_0=1$). To simplify further the above expression, we evaluate explicitly the term
$$
e_\eta\tra{\sigma_i\sigma_j\sigma_\eta}=2\delta_{ij}+2i\epsilon_{ijk}e_k.
$$
Since $\mathcal{S}$ is symmetric, we have $\mathcal{S}_{ij}\epsilon_{ijk}=0$, and \eqref{T-one} simplifies to
$$
T=\mathcal{S}_{ii}-\mathcal{S}_{ij}e_ie_j.
$$
Defining $S = ({\cal S})_{ij}$ as the $3\times3$ `spatial' sub-block of ${\cal S}$, we can write
$$
T=\tr{S}-\vec e^{\sf T} S\vec e.
$$
Since $|\vec e|=1$, the minimization of such expression yields Eq.~(\ref{dggen}),
\begin{equation}\label{dggen}
{\cal D}_G(\rho_{AB}) = \text{tr} (S) - \lambda_{\max}(S)\,,\quad \mbox{with $S=\text{tr}_B[\vec{v} \vec{v}^{\sf T}]$}\,.
\end{equation}
Recall that $S=\trb{\vec v\vec v^{\sf T}}$, with $\vec{v}=\tra{\rho_{AB}\vec{\sigma}}$).

Equation (\ref{dggen}) encompasses the known formulas for two-qubit \cite{dakic,luofu} and qubit-qudit states \cite{sai,gharibian,q}, but is valid as well for states of a qubit and a qumode, for which $d=\infty$. In the latter case, it can be convenient to adopt a hybrid Hilbert-space/phase-space picture to describe qubit-oscillator states $\rho_{AB}$ \cite{klenner,tufophd}. Let $b$ ($b^\dagger$) denote the annihilation (creation) operator for the qumode, with $[b,b^\dagger]=1$, and let $D(\beta) =\exp(\beta b^\dagger - \beta^\ast b)$ denote the corresponding Weyl displacement operator, with $\beta \in \mathbbm{C}$. We can define the {\it characteristic vector} associated to the state $\rho_{AB}$ as $\boldsymbol{\chi}(\beta) \equiv \big(\chi_0(\beta), \vec{\chi}(\beta)\big) = \text{tr}\big[\rho_{AB} \boldsymbol{\sigma} D(\beta)\big] = \text{tr}_B\big[{\bf v} D(\beta)\big]$, where the zero-th component $\chi_0(\beta)=\text{tr}_B \big[\rho_B D(\beta)\big]$ is the conventional characteristic function of the oscillator \cite{quantumoptics}, describing its marginal state in phase space. The matrix $S$ appearing in Eq.~(\ref{dggen}) is then $S=\pi^{-1} \int {\rm d}^2 \beta  \vec{\chi}(\beta) \vec{\chi}^\dagger(\beta)$. Similar representations can be provided by employing e.g.~the Wigner distribution to describe the oscillator \cite{klenner,lukt,tufophd}.

\subsection{Explicit calculation for the states $\boldsymbol{\rho_{AB}}$}

To calculate $\mathcal{D}_G(\rho_{AB})$, we need the spatial components of the vector ${\bf v}$ (see previous subsection). These are
\begin{align}
&v_1=r D(\beta)\rho_B^0 D(\beta)+r^* D^\dagger(\beta)\rho_B^0 D^\dagger(\beta),\label{v1}\\
&v_2=i r D(\beta)\rho_B^0 D(\beta)- i r^* D^\dagger(\beta)\rho_B^0 D^\dagger(\beta),\label{v2}\\
&v_3=p D(\beta)\rho_B^0 D^\dagger(\beta)-(1-p)D^\dagger(\beta)\rho_B^0 D(\beta).\label{v3}
\end{align}
The matrix $S$ can now be calculated according to $S_{ij}=\trb{v_iv_j}$. We are interested in the regime of large displacements. Suppose that $|\beta|$ is large compared to the phase space extension of the initial oscillator state $\rho^0_B$. Then the matrix $S$ converges to
\begin{equation}S \underset{|\beta|\rightarrow \infty}{\longrightarrow}
\left(\begin{array}{ccc}
2|r|^2&0&0\\
0&2|r|^2&0\\
0&0&2p(1-p)+1
\end{array}\right)\trb{(\rho^0_B)^2}.\label{slingua}
\end{equation}
From the above, as well as the condition $|r|^2\leq p(1-p)$, we can directly see that $\lambda_{\text{max}}(S)=[2p(1-p)+1] \trb{(\rho_B^0)^2}$. Hence, the geometric discord of the state $\rho_{AB}$, for the case of large displacements, is given by Eq.~(\ref{dg}), $\mathcal{D}_G(\rho_{AB}) \underset{|\beta|\rightarrow \infty}{\longrightarrow} 4|r|^2\trb{(\rho_B^0)^2}$. Note that $\mathcal{D}_G(\rho_{AB})\leq\trb{(\rho_B^0)^2}$, i.e.,  the geometric discord is smaller than the purity of the initial state of the oscillator.

To see that Eq.~(\ref{slingua}) is correct in the limit of large displacements, we can proceed as follows.
When we calculate $\trb{v_iv_j}$, we get terms of the form $\trb{\rho_BD_1D_2\rho_B D_3D_4}$, where each one of the $D_j$'s can be either $D(\beta)$ or $D^\dagger(\beta)$. In the limit of large $|\beta|$, however, only those that evaluate to $\trb{(\rho^0_B)^2}$ survive. All the others can be shown to be negligible by using the fact that $\left|\trb{\rho_BD(\alpha)}\right|\rightarrow 0$ for $|\alpha|\gg|\beta|$ (see also Lemma 2 in the previous Appendix).
For example, using Glauber's $P$-representation we have $\rho_B=\int{\rm d}^2\alpha P(\alpha)\kebra{\alpha}{\alpha}$, and we can see that (denoting $D(\beta)\equiv D$), for instance,
\begin{align}
\abs{\trb{\rho_B (D^\dagger)^2\rho_B D^2}}&\leq \int{\rm d}^2\alpha{\rm d}^2\alpha'\;|P(\alpha)P(\alpha')||\bra{\alpha'} D^2\ket{\alpha}|^2=\nonumber\\
&=\int{\rm d}^2\alpha{\rm d}^2\alpha'\;|P(\alpha)P(\alpha')|{\rm e}^{-|2\beta+\alpha-\alpha'|^2} \nonumber \\
&\underset{|\beta|\rightarrow \infty}{\longrightarrow} 0\,,
\end{align}
and so on. The above and all the other terms can be evaluated explicitly e.g.~in case $\rho_B^0$ is a Gaussian state, but this restriction is not crucial for the validity of Eq.~(\ref{dg}).

\section{Digitalization of the qubit-oscillator states $\boldsymbol{\rho_{AB}}$}\label{secdigapp}

\noindent {\bf Theorem 3.} {\it For any $\epsilon>0$, it is possible to find a $\beta \in \mathbbm{C}$ and a local operation on Bob, $\sum_jO_j\rho_\textrm{{AB}}O_j^\dagger$, with $\sum_j O_j^\dagger O_j=\mathbbm{1}$, such that the state $\rho_{AB}$ [Eq.~\eqref{states}] is converted into the `digitalized' two-qubit state
\begin{align}
\tilde{\rho}_{AB}^{\rm dig}&=p\kebra{e}{e}\otimes \kebra{\tilde{e}}{\tilde{e}}+(1-p)\kebra{g}{g}\otimes \kebra{\tilde{g}}{\tilde{g}}\nonumber \\ &+r\kebra{e}{g}\otimes \kebra{\tilde{e}}{\tilde{g}}+r^*\kebra{g}{e}\otimes \kebra{\tilde{g}}{\tilde{e}},\label{digitalized}
\end{align}
with probability $P\geq1-\epsilon$ and fidelity ${\cal F}>1-\epsilon$.}

\smallskip

\noindent {\it Proof.}
Recall $\rho^0_B=\sum_n s_n \kebra{\psi_n}{\psi_n}$.
Let us fix $\epsilon_N=\sum_{n=N+1}^\infty s_n$; clearly $\lim_{N\to\infty}\epsilon_N=0$. Let us now consider the two subspaces $\mathcal H_{1,2}$ spanned by the bases $\mathcal B_1=\{D(\beta)\ket{\psi_1},...,D(\beta)\ket{\psi_N}\}$ and $\mathcal B_2=\{D(\beta)^\dagger\ket{\psi_1},...,D(\beta)^\dagger\ket{\psi_N}\}$. The bases $\mathcal B_1$ and $\mathcal B_2$ may overlap with each other; nevertheless, in the limit $|\beta|\to \infty$, $\mathcal B=\{\mathcal B_1,\mathcal B_2\}$ becomes an orthonormal set. Let us then choose $|\beta|$ large enough such that this orthonormality is verified for all practical purposes. We denote by $\mathcal H_{12}=\mathcal H_1\oplus\mathcal H_2$, the subspace spanned by the basis $\mathcal B$. Let us consider the local operation on Bob, corresponding to
$$
O_j=\kebra{\tilde{e}}{\psi_j}D^\dagger(\beta)+\kebra{\tilde{g}}{\psi_j}D(\beta),
$$
where  $j=0,\ldots,N$. To see that those operators are quasi-complete, we evaluate
\begin{align}
\sum_j O^\dagger_j O_j&=D(\beta)\sum_j^N\kebra{\psi_j}{\psi_j}D^\dagger(\beta)+D^\dagger(\beta)\sum_j^N\kebra{\psi_j}{\psi_j}D(\beta)\nonumber
\\ &=\mathcal P_{12},
\end{align}
where $\mathcal P_{12}$ is the projection on the subspace $\mathcal H_{12}$. Let us in general denote the complete operation with $\{O_{j=1}^N,O_\perp\}$. It is irrelevant how specifically we complete the operation on the remainder of Bob's Hilbert space; we can just assume that the fidelity of the output state is zero if we go outside $\mathcal H_{12}$ (this happens with probability $<\epsilon_N$).
Now we can easily see that
$$
	\lim_{\beta\to\infty} \sum_{j=0}^NO_j\rho_\textrm{AB}O_j^\dagger=\sum_{j=0}^N s_j\tilde{\rho}^{\rm dig}_{AB}
$$
so that the output of the complete operation is some state
$$
	 \tilde{\rho}^\textrm{out}_{AB}=(1-\epsilon_N)\rho^{\rm dig}_{AB}+\epsilon_N\tilde{\rho}^\perp_{AB},
$$
where $\tilde{\rho}^\perp_{AB}$ is some state orthogonal to $\tilde{\rho}^{\rm dig}_{AB}$. It is clear that, if we choose $\epsilon_N$ small enough, this state $\tilde{\rho}^\textrm{out}_{AB}$ has fidelity ${\cal F}>1-\epsilon$ with the target state $\tilde{\rho}^{\rm dig}_{AB}$ of Eq.~(\ref{digitalized}). \hfill $\blacksquare$


\begin{thebibliography}{299}
\bibitem{footnotemixedqub}
A similar dependence on the resource state mixedness has been reported for the two-qubit version of remote state preparation \cite{chaves}, whose figure of merit has been operationally linked to the geometric discord \cite{dakicrsp}.
\bibitem{noteex}For example, if $\rho_B^{0}$ is a Gaussian thermal state, then $\ket{\psi_m}=\ket m$ are the Fock states, and Bob needs to be able to discriminate photon numbers going from $\ket0$ to $\ket N$, as well as the more complicated states given by $[D(2\beta)\pm D^\dagger(2\beta)]\ket m$.
\end{thebibliography}
\end{document}